\documentstyle[12pt]{article}
\begin{document}

\begin{flushright}
\begin{tabular}{l}
 JHU-TIPAC-95012 \\
 MADPH-95-877 \\
CPP-95-3 \\
DOE-ER40757-063 \\
UCD-95-08 \\
MSUHEP-50305 \\
IFUSP/P-1141\\
hep-ph/9504426\\
April 1995
\end{tabular}
\end{flushright}

\bigskip

\begin{center}
{\large\bf LHC Analysis of the Strongly Interacting $WW$ System:\\
Gold-Plated Modes}\\[.25in]
J. Bagger,${}^{(a)}$ V. Barger,${}^{(b)}$ K. Cheung,${}^{(c)}$
J. Gunion,${}^{(d)}$ T. Han,${}^{(d)}$\\[.1in]
G. A. Ladinsky,${}^{(e)}$ R. Rosenfeld${}^{(f)}$ and
C.--P. Yuan${}^{(e)}$

\medskip

\parbox{5.5in}{
\begin{itemize}
\item[(a)]{\it Department of Physics and Astronomy, Johns Hopkins
University,
\\  Baltimore, MD  21218}
\item[(b)]{\it Department of Physics, University of Wisconsin,
Madison, WI  53706}
\item[(c)]{\it  Center for Particle Physics, University of Texas at Austin,
Austin,
\\ TX  78712}
\item[(d)]{\it Davis Institute for High Energy Physics,
Department of Physics,
\\ University of California at Davis, Davis, CA  95616}
\item[(e)]{\it Department of Physics and Astronomy, Michigan State University,
\\ East Lansing, MI 48824}
\item[(f)]{\it Instituto de Fisica, Universidade de Sao Paulo,
P.O.Box 20516,
\\ Sao Paulo, Brazil}
\end{itemize}}
\end{center}

\vskip .3in

\centerline{\bf Abstract}
\smallskip

We study the gold-plated purely leptonic signal and background rates
at the LHC for
the $ZZ,$ $W^{+}W^-,$ $W^\pm Z$ and $W^\pm W^\pm$ final states
associated with strongly interacting electroweak symmetry breaking.
We work at an energy of $\sqrt s = 14$ TeV, and develop
a combination of back-to-back leptonic, central-jet-vetoing and
forward-jet-tagging cuts
that suppresses the Standard-Model backgrounds.
We find that the LHC with an annual luminosity of 100 fb$^{-1}$
will achieve a reasonably good sensitivity to the physics of
strongly interacting electroweak symmetry breaking.

\thispagestyle{empty}

\newpage
\section{Introduction}

In a recent paper \cite{baggeretali}, we studied the signals and
backgrounds for strongly interacting electroweak symmetry
breaking \cite{lqtetal,changail}  at hadron supercolliders.
We analyzed $WW$ scattering in a series
of models that are consistent with present data.  (In this paper,
the symbol $W$ denotes either $W^\pm$ or $Z$, unless specified
otherwise.)  We concentrated on the gold-plated purely leptonic decays
of the final-state $W$'s, {\it e.g.} $W^\pm \rightarrow \ell^\pm \nu$
and $Z \rightarrow  \ell^+ \ell^-$ ($\ell=e,\mu$) to avoid large QCD
backgrounds.  We developed a set of techniques,
including back-to-back leptonic,  forward-jet-tagging and central-jet-vetoing
cuts,  that proved sufficient to isolate the
strongly interacting signal from the Standard-Model background.  For
each model, we found a statistically significant sensitivity in at
least one channel.

In our previous work, our cuts were optimized for the SSC environment.
Nevertheless, we found significant signals at an LHC energy of $\sqrt
s =16$ TeV, assuming an annual luminosity of 100 fb$^{-1}$.  In this
paper, we refine our LHC analysis using the currently-planned energy of
$\sqrt s=14$ TeV.  As before, we consider seven different models to
represent the possible physics associated with a strongly interacting
electroweak symmetry breaking sector:

\begin{enumerate}
\item
SM:  The Standard Model with $M_H = 1~{\rm TeV}$;

\item
Scalar:  A nonlinearly-realized chiral model with a spin-zero,
isospin-zero resonance of mass $M_S = 1~{\rm TeV}$ and width
$\Gamma_S = 350~{\rm GeV}$;

\item
O($2N$):  A large-$N$ model \cite{einhorn}
that gives rise to an amplitude
with a pole at $s=[M-i\Gamma/2]^2$, where $M=0.8~{\rm TeV}$
and $\Gamma=600~{\rm GeV}$ for $N=2$ and high-energy
cutoff $\Lambda = 3$ TeV;

\item
Vector:  A nonlinearly-realized chiral model with a spin-one,
isospin-one resonance;  we choose the mass-width combinations
$(M_V,\Gamma_V)=$ (1 TeV, 5.7 GeV) and (2.5 TeV, 520 GeV);

\item
LET-CG:  A non-resonant nonlinearly-realized chiral model based on the
Low Energy Theorems (LET) for Goldstone boson scattering.  The scattering
amplitudes are unitarized by cutting off the tree-level partial waves
when they reach the unitarity bound \cite{changail};

\item
LET-K:  The same model as above, except that the scattering amplitudes
are unitarized by the K-matrix technique;

\item
Delay-K:  A non-resonant nonlinearly-realized chiral model in which
the scattering amplitudes are computed to second order in the energy
expansion.  The counterterms are chosen to delay the unitarity violation
to energies beyond $2~{\rm TeV}$ \cite{bdv}.
K-matrix unitarization is used to
ensure unitarity beyond this point.

\end{enumerate}

\noindent
These models are described fully in Ref.~\cite{baggeretali}.

For each of these models, we consider the following final-state modes,
\[
\begin{array}{c}
ZZ \rightarrow \ell^+\ell^-\ell^+\ell^-, \qquad ZZ \rightarrow
\ell^+\ell^-\nu\overline\nu, \qquad
W^\pm Z \rightarrow \ell^\pm\nu\ell^+\ell^-,  \\
W^+W^- \rightarrow \ell^+\nu\ell^-\overline\nu,
\qquad W^\pm W^\pm \rightarrow \ell^\pm\nu\ell^\pm\nu .
\end{array}
\]
We include the process $ZZ \rightarrow \ell^+ \ell^- \nu \bar\nu$ \cite{silver}
to enhance the statistical significance of the $ZZ$ channel.  For
the models with vector resonances, we also compute $q \bar q'
\rightarrow W^* \rightarrow V \rightarrow WW$ \cite{bhr,ck}.
This $q \bar q$
annihilation process is more important at LHC than SSC energies.

Our background analysis includes the ``irreducible EW'' processes,
$qq'  \rightarrow qq'  W_TW_T$  $(W_TW_L)$, and the ``continuum WW''
processes, $q\bar q' \rightarrow WW +\, {\rm QCD ~jets}$.  We re-evaluate
the continuum contribution to include all ${\cal O}(\alpha_s)$ corrections
\cite{ohnemus}, which are important in the kinematical region of interest.
We also update the top quark backgrounds, $t \bar t + {\rm  jets}$ and
$t\bar t W +\ {\rm  jets},$ for $m_t = 175$ GeV.  Finally, we include a
new detector-dependent background, $W^+Z \rightarrow \ell^+ \ell^+ X$.
This is a background to the  $W^+W^+$ final state \cite{atlas,ck}.
In our numerical calculations, we use the recent parton distribution
functions MRS Set-A~\cite{mrsa}.  Other recent work along similar lines
can be found in Refs.~\cite{ck,atlas}.

\section{$WW$ Fusion}

At hadron supercolliders, the physics of strongly interacting electroweak
symmetry breaking can be studied through the fusion of two
longitudinally-polarized
$W$'s.  These particles interact and then rescatter into the final state.  The
final-state $W$'s then decay to leptons, giving events that are characterized
by several distinct features which can be used to separate the signal from the
background \cite{wpwp,tagveto}:

\begin{enumerate}
\item[(i)]  The signal events contain very energetic leptons in the central
(low rapidity) region.  The leptons from $W^+ W^-$ and $W^\pm W^\pm$
are very back-to-back, so $\cos\phi_{\ell
\ell}$ is near $-1$ and $\Delta p_T(\ell\ell) \equiv |{\bf p_T}(\ell_1)
- {\bf p_T}(\ell_2)|$ is large ($\phi_{\ell\ell}$ is the azimuthal
angle between a lepton $\ell_1$ from one $W_L$ and a lepton $\ell_2$
from the other).  The invariant mass $M(\ell\ell)^2 = (p_{\ell_1}+
p_{\ell_2})^2$ is large as well.

\item[(ii)]  The signal events have low hadronic jet activity in the central
region;

\item[(iii)] The signal events also contain two highly energetic, low-$p_T$,
high-rapidity ``spectator'' jets.

\end{enumerate}

In our previous paper, we found that stringent leptonic cuts, when
combined with a veto of events with hard central jets and a tag of
energetic spectator jets, were sufficient to dramatically suppress the
large backgrounds.  In particular, we found the most difficult irreducible
electroweak backgrounds to be those from $W_TW_T$ and $W_TW_L$
final states at ${\cal O}(\alpha^4)$.
These backgrounds have spectator jets that are more central and leptons that
are less back-to-back than those from  the $W_LW_L$ signal.
The central-jet veto plus $\cos\phi_{\ell\ell} < -0.8$
and large-$\Delta p_T(\ell\ell)$ cuts proved to
be effective against these important backgrounds.
After carefully examining the kinematics of the signal and backgrounds,
our re-optimized cuts are summarized in Table~\ref{cutstable}.
%\footnote{
Note that the transverse mass variables  are  defined as
$M_T^2(ZZ) = [ \sqrt{M^2_Z + p_T^2(\ell \ell)} +
\sqrt{M^2_Z + |{p_T^{miss}}|^2 } ]^2
- [ \vec{p}_T(\ell \ell) + \vec{ p}_T^{miss} ]^2,$ and
$M_T^2(WZ) = [\sqrt{M^2(\ell \ell \ell) + p_T^2(\ell \ell \ell)}
+ |p_T^{miss}| ]^2 - [ \vec{p}_T(\ell \ell \ell) + \vec{ p}_T^{miss} ]^2 .$

\begin{table}
\centering

\caption{\label{cutstable}
Leptonic, single-jet-tagging and central-jet-vetoing
cuts for generic $W_LW_L$ fusion processes
at the LHC energy,  by final-state mode.}
\bigskip

\begin{tabular}{lcc}
\hline\hline
$Z Z(4\ell)$ & Leptonic Cuts & Jet Cuts \\
\hline
 \ &
 $\vert  y({\ell}) \vert  < 2.5 $  &
 $E(j_{tag}) > 0.8~{\rm TeV}$   \\
 \ &
 $p_T(\ell) > 40~{\rm GeV}$  &
 $3.0 < \vert y(j_{tag}) \vert < 5.0$   \\
 \ &
 $p_T(Z) > {1\over4} \sqrt{M^2({ZZ}) - 4 M^2_Z}$  &
 $p_T(j_{tag}) > 40~{\rm GeV}$   \\
 \ &
 $M({ZZ}) > 500~{\rm GeV}$  & No Veto \\
\hline
$Z Z(\ell\ell\nu\nu)$ & Leptonic Cuts & Jet Cuts \\
\hline
 \ &
 $\vert  y({\ell}) \vert  < 2.5 $  &
 $E(j_{tag}) > 0.8~{\rm TeV}$   \\
 \ &
 $p_T(\ell) > 40~{\rm GeV}$  &
 $3.0 < \vert y(j_{tag}) \vert < 5.0$   \\
 \ &
 $p_T^{\rm miss} > 250~{\rm GeV}$ &
 $p_T(j_{tag}) > 40~{\rm GeV}$   \\
 \ &
 $M_T(ZZ)
%\footnote{define}
> 500~{\rm GeV}$  &
 $p_T(j_{veto}) > 60~{\rm GeV}$ \\
 \ &
 $p_T{(\ell\ell)}>M_T(ZZ)/4$ &
 $ \vert  y(j_{veto}) \vert  < 3.0$  \\
\hline
$W^+W^-$ & Leptonic Cuts & Jet Cuts \\
\hline
 \ &
 $\vert y({\ell}) \vert < 2.0 $  &
 $E(j_{tag}) > 0.8~{\rm TeV}$  \\
 \ &
 $p_T(\ell) > 100~{\rm GeV}$  &
 $3.0 < \vert y(j_{tag}) \vert < 5.0$  \\
 \ &
 $\Delta p_T({\ell\ell}) > 440~{\rm GeV}$  &
 $p_T(j_{tag}) > 40\ {\rm GeV}$  \\
 \ &
 $\cos\phi_{\ell\ell} < -0.8$  &
 $p_T(j_{veto}) > 30~{\rm GeV}$  \\
 \ &
 $M({\ell\ell}) > 250~{\rm GeV}$  &
 $ \vert  y(j_{veto}) \vert  < 3.0$  \\
\hline
$W^\pm Z$ & Leptonic Cuts & Jet Cuts\\
\hline
 \ &
 $\vert  y({\ell}) \vert  < 2.5 $  &
 $E(j_{tag}) > 0.8~{\rm TeV}$  \\
 \ &
 $p_T(\ell) > 40~{\rm GeV}$  &
 $3.0 < \vert y(j_{tag}) \vert < 5.0$  \\
 \ &
 $ p_{T}^{\rm miss} >  50~{\rm GeV}$  &
 $p_T(j_{tag}) > 40~{\rm GeV}$   \\
 \ &
 $p_T(Z) > {1\over4} M_T(WZ) $ &
 $p_T(j_{veto}) > 60~{\rm GeV}$  \\
 \ &
 $M_T(WZ)
%\footnote{$
%M_T^2 = \bigg(\sqrt{M^2(\ell \ell \ell) + p_T^2(\ell \ell \ell)}
%+ |p_T^{miss}|\bigg)^2 -
%( {\bf p}_T(\ell \ell \ell) + {\bf  p}_T^{miss} )^2 .$}
> 500\ {\rm GeV}$ &
 $ \vert  y(j_{veto}) \vert  < 3.0$  \\
\hline
$W^\pm W^\pm $ & Leptonic Cuts & Jet Cuts \\
\hline
 \ &
 $\vert  y({\ell}) \vert  < 2.0 $   &
   \\
 \ &
 $p_T(\ell) > 70~{\rm GeV}$   &
  $3.0 < \vert y(j_{tag}) \vert < 5.0$  \\
 \ &
 $\Delta p_T(\ell\ell) > 200~{\rm GeV}$   &
  $p_T(j_{tag}) > 40~{\rm GeV}$   \\
 \ &
 $\cos\phi_{\ell\ell} < -0.8$  &
 $p_T(j_{veto}) > 60~{\rm GeV}$ \\
 \ &
 $M(\ell\ell) > 250~{\rm GeV}$ &
 $\vert y(j_{veto}) \vert < 3.0$  \\
\hline\hline
\end{tabular}
\end{table}

As before \cite{baggeretali},
we use the Effective-$W$ Approximation (EWA) and the Equivalence
Theorem (ET) to compute the cross sections for the $W_LW_L \to W_LW_L$ signals.
We do this by first computing the cross sections ignoring all jet observables,
implementing the lepton cuts by decaying the final-state $W_L$'s according
to their appropriate angular distributions.  We then approximate the
jet-tagging
and vetoing cuts by multiplying our cross sections by tagging and/or vetoing
efficiencies determined from an exact SM calculation with a 1 TeV Higgs boson
(see Table~\ref{smlhctable}).  This procedure is accurate because the
kinematics
of the jets in the signal events are determined by the initial-state $W_L$'s
and  are insensitive to  the strong $W_L W_L$ scattering dynamics in the
TeV region.

Since the jet-tagging and vetoing efficiencies are so important for our
results, we will review our procedure, focusing on the case of the SM
with an ordinary Higgs boson.  We start with an exact calculation using
the full SM matrix elements  of  ${\cal O}(\alpha^4)$
for $q\overline q' \rightarrow q \overline q'
WW \rightarrow q\overline q' + 4~{\rm leptons}.$   We include all
final-state polarizations, and consider two cases: a heavy 1 TeV Higgs
boson, and a light 0.1 TeV Higgs particle.

We evaluate the cross section for the SM with a light Higgs boson because
it should accurately represent the perturbative irreducible backgrounds
from the $qq'  \rightarrow qq' W_TW_T$ $(W_TW_L)$ processes.  This background
is inevitable because of our inability to experimentally determine the
polarizations of the final-state $W$'s on an event-by-event basis.
We emphasize that the rate and kinematical distributions of $W_TW_T
(W_TW_L)$  events  are quite insensitive to the particular model adopted
for strongly interacting $W_LW_L$ scattering.

Given these considerations, we define the $W_LW_L$ signal to be the total
enhancement over the SM prediction for a light 0.1 TeV Higgs boson.  For
the case of 1 TeV Higgs boson, this implies
\begin{equation}
\sigma(W_LW_L~{\rm signal}) \equiv \sigma({\rm SM}~M_H=1~{\rm TeV}) -
     \sigma({\rm SM}~M_H=0.1~{\rm TeV})\,.
\label{signaldef}
\end{equation}
The jet-cut efficiencies are determined from this signal definition.  We
show in Table~\ref{smlhctable} the SM results for the cross sections, after
imposing the cuts in Table~\ref{cutstable}.  We also show the SM backgrounds
as well as the resulting jet-tagging and vetoing efficiencies for a 1 TeV
Higgs boson signal.  We use these efficiencies for all the models of strong
electroweak symmetry breaking.

\begin{table}
\centering
\caption{\label{smlhctable}
Standard-Model cross sections (in fb) for electroweak processes $q\bar q'
\to q\bar q' WW$, for $M_H=1$ TeV and $0.1$ TeV.  Also given are the
cross sections for continuum $WW$
production at ${\cal O}(\alpha^2 \alpha_s)$, and other backgrounds, with
$\protect\sqrt{s} =  14$ TeV and $m_t = 175$ GeV.}
\bigskip
\begin{tabular}{cccc}
\hline\hline
$Z Z(4\ell) $ & Leptonic Cuts Only & $+$ Veto Only / Veto Eff. &
$+$ Veto $+$ Tag / Tag Eff. \\
\hline
 EW($M_H=1.0~{\rm TeV}$) & 0.12  & - & 0.045 / 39\% \\
 EW($M_H=0.1~{\rm TeV}$) & 0.019  & - & 0.004 / 19\% \\
  Continuum $ZZ$           & 0.42  & - & 0.003 / 0.6\% \\
 $Z_LZ_L$ signal     & 0.098  & - & 0.041 / 43\% \\
\hline
$Z Z(2\ell2\nu) $ & Leptonic Cuts Only & $+$ Veto Only / Veto Eff. &
$+$ Veto $+$ Tag / Tag Eff. \\
\hline
 EW($M_H=1.0~{\rm TeV}$) & 0.69  & 0.30 / 43\%  & 0.16 / 54\% \\
 EW($M_H=0.1~{\rm TeV}$) & 0.11  & 0.014 / 13\%  & 0.006 / 38\% \\
  Continuum $ZZ$          & 2.2  & 1.7 / 75\%  & 0.012 / 0.7\% \\
 $Z_LZ_L$ signal     & 0.59  & 0.29 / 49\%  & 0.16 / 55\% \\
\hline
$W^+W^-$ & Leptonic Cuts Only & $+$ Veto Only / Veto Eff. &
$+$ Veto $+$ Tag / Tag Eff. \\
\hline
 EW($M_H=1.0~{\rm TeV}$) & 1.1 & 0.33 / 30\%  & 0.20 / 59\% \\
 EW($M_H=0.1~{\rm TeV}$) & 0.32 & 0.039 / 12\%  & 0.016 / 40\% \\
 Continuum $W^+W^-$          & 6.8 & 3.5 / 51\%  & 0.041 / 1.2\% \\
 $t\overline t+{\rm jet}$        & 59 & 0.88 / 1.5\% & 0.067 / 7.7\% \\
 $W_LW_L$ signal     & 0.80 & 0.29 / 37\%  & 0.18 / 61\% \\
\hline
$W^\pm Z$ & Leptonic Cuts Only & $+$ Veto Only / Veto Eff. &
$+$ Veto $+$ Tag / Tag Eff. \\
\hline
 EW($M_H=1.0~{\rm TeV}$) & 0.32 & 0.07 / 22\% & 0.032 / 46\% \\
 EW($M_H=0.1~{\rm TeV}$) & 0.25 & 0.043 / 17\% & 0.018 / 42\% \\
 Continuum  $W^\pm Z$  & 3.8 & 2.2 / 56\% & 0.03 / 1.4\% \\
 $Z t\overline t+{\rm jet}$        & 0.42 & 0.008 / 2.0\% & 0.001 / 16\% \\
 $W_LZ_L$ signal     & 0.073 & 0.027 / 37\%  & 0.014 / 52\% \\
\hline
$W^\pm W^\pm $ & Leptonic Cuts Only & $+$ Veto Only / Veto Eff. &
$+$ Veto $+$ Tag / Tag Eff. \\
\hline
 EW($M_H=1.0~{\rm TeV}$) & 0.66 & 0.15 / 23\% & 0.099 / 66\% \\
 EW($M_H=0.1~{\rm TeV}$) & 0.45 & 0.057 / 13\% & 0.034 / 60\% \\
 $g$-exchange        & 0.15 & 0.009 / 6.0\% & 0.001/ 7.7\% \\
 $Wt\overline t$         & 0.42 & 0.012 / 3.0\% & 0.001 / 13\% \\
 Continuum  $W^\pm Z$  & 0.15 & 0.10/ 65\% & 0.001 / 1.4\% \\
 $W_LW_L$ signal     & 0.22 & 0.093 / 43\%  & 0.066 / 70\% \\
\hline\hline
\end{tabular}
\end{table}

We summarize some important details regarding our background computations
below.

\begin{enumerate}

\item[(a)] We calculate the background processes for continuum $WW$
production including full 1-loop QCD corrections at
${\cal O}(\alpha^2 \alpha_s)$ \cite{ohnemus}.
This procedure incorporates large QCD corrections
in background rates (as large as 70\% for the inclusive cross section),
and also allows us to reliably
determine the leading-jet kinematics in both the central and
forward regions.

\item[(b)] As noted earlier, we enhance the utility of the $ZZ$ final states
by considering the $ZZ \to 2\ell 2\nu$ channel \cite{silver}
as well as the cleaner, but
lower rate, $ZZ \to 4\ell$ mode.  For the $ZZ \to 2\ell 2\nu$ channel
there is a detector-dependent background from $Z+$QCD jets, where
$Z \to 2\ell$ and
the jets are missing along the beampipe or are mismeasured by
the calorimeter, resulting in significant missing $p_T$. Recent
studies including detector simulations at the LHC \cite{atlas,cms}
show that this background can be suppressed below that from continuum
$ZZ$ production by requiring $p_T^{\rm miss}>250$ GeV.  Thus, we have
imposed this cut and neglected the $2\ell+$ QCD jets background.

\item[(c)] Obviously, $t\bar t$ production overwhelms the $W_L^+W_L^-$
signal in total rate at the LHC energy.  Nonetheless, our combination
of stringent leptonic cuts, together with the central-jet veto and
forward-jet tag, reduces the background to a manageable level.

\item[(d)] The 1 TeV Higgs boson contribution to the SM $W^\pm Z$ process
(via $t$-channel exchange)  is hardly significant, so we have optimized our
cuts in this channel for a vector resonance signal with $M_V \sim 1$ TeV
and $\Gamma_V \sim$ 5.7 GeV.

\item[(e)] Since $q\overline q$ annihilation to $W^\pm W^\pm $ is not
possible, the leading QCD-related background comes from diagrams in
which a gluon is exchanged between two colliding quarks which then
radiate the two like-sign $W$'s.  We denote this by  ``$g$-exchange''
\cite{ewa,gluon,wpwp}.
We find that our cuts effectively remove this background despite the
kinematic similarity of its final state to the $WW$ scattering reaction
of interest.  This type of process also contributes to all the other
channels, but is of higher order, ${\cal O}(\alpha^2 \alpha_s^2)$,
than the continuum $WW$ processes that we include,
${\cal O}(\alpha^2 \alpha_s)$.

\item[(f)]  Because of the finite coverage of the EM calorimeter,
$|y(\ell)| \sim 3$ \cite{atlas,cms}, one must consider the background
to $W^\pm W^\pm $ from $W^\pm Z$ when the extra
$\ell^\mp$ from $Z$ decay is not detected \cite{atlas,ck}.
As we see from Table~\ref{smlhctable}, this continuum
$W^\pm Z$ background is as large as the $W^\pm W^\pm $ signal after
leptonic and central-jet-vetoing cuts.  The most efficient means
for removing this background is to include a forward-jet-tagging cut.

In fact, the  $W^\pm Z$ process may also be a background to $W^+ W^-$
if  the  $\ell^\pm$ from $Z$ decay is not detected.  However, this is
far less important than the $t \bar t$ background; therefore it is not
surprising that it is eliminated by the stringent leptonic and
jet-tagging cuts in the $W^+ W^-$ channel.

\item[(g)] Of course, we have implicitly assumed
sufficiently good isolation for
identified charged leptons that faked backgrounds from $c,b$-quark
semileptonic decays are not significant, see Ref. \cite{baggeretali}.

\end{enumerate}

%\begin{table}[h]
%\begin{table}[b]
\begin{table}[t]
\centering
\caption{\label{lhcrates}
Event rates per LHC-year for $W_LW_L$ fusion signals from the different models,
together with backgrounds, assuming $\protect\sqrt s=14~{\rm TeV}$, an annual
luminosity
of $100 ~{\rm fb}^{-1}$, and $m_t=175~{\rm GeV}$.  Cuts are listed in
Table~\protect\ref{cutstable}.  Jet-vetoing and tagging efficiencies are listed
in Table~\protect\ref{smlhctable}.
The $W^\pm Z(M_T^{cut})$ row refers to the $W^\pm Z$ events
with 0.8 $< M_T(WZ)<$ 1.1 TeV, optimized to search for
a 1 TeV isovector signal.}
\bigskip

\begin{tabular}{l|ccccccccc}
\hline\hline
& Bkgd. & SM & Scalar & O($2N$) & Vec~1.0 & Vec~2.5 & LET-CG &
   LET-K & Delay-K  \\
\hline
$Z Z(4\ell)$  & 0.7 & 9 & 4.6 & 4.0 & 1.4 & 1.3 & 1.5 & 1.4 & 1.1  \\
 \hline
$Z Z(2\ell2\nu)$ & 1.8 & 29 & 17 & 14 & 4.7 & 4.4 & 5.0 & 4.5 & 3.6  \\
 \hline
$W^+W^-$ & 12 & 27 & 18 & 13 & 6.2 & 5.5 & 5.8 & 4.6 & 3.9 \\
  \hline
$W^\pm Z$ & 4.9 & 1.2 & 1.5 & 1.2 & 4.5 & 3.3 & 3.2 & 3.0 & 2.9 \\
$W^\pm Z(M_T^{cut})$ & 0.82 &  &  &  & 2.3 &  &  &  &  \\
  \hline
$W^\pm W^\pm $ & 3.7 & 5.6 & 7.0 & 5.8 & 12 & 11 & 13 & 13 & 8.4 \\
\hline\hline
\end{tabular}
\end{table}

In Table~\ref{lhcrates} and Fig.~\ref{sewsfigi}  we present our results
for $W_LW_L$ fusion signals versus the SM backgrounds for the different
models described above:

\begin{itemize}

\item The isoscalar models (SM, Scalar, O($2N$)) give rise to substantial
signals over backgrounds in the $ZZ \to 4\ell$ and $ZZ \to 2\ell2\nu$ channels.
Especially encouraging is the signal rate for the  $2\ell 2\nu$ mode.  The
$W^+ W^-$ channel also exhibits some sensitivity to these models; the actual
sensitivity is probably somewhat greater since the distribution in the mass
variable $M(\ell\ell)$ peaks broadly around half the mass of the scalar
resonance.

\item The isovector models (Vec~1.0, Vec~2.5) yield a continuum  event
excess in the $W^\pm W^\pm $ channel, and, to a  lesser extent, in the
$W^\pm Z$ channel, where the signal rates are rather low and the background
level remains difficult.  Nevertheless,  as seen in the $M_T$-distribution
for the $W^\pm Z$ channel,
it might be possible to search for a signal peak
if $M_V \sim 1$ TeV.  As indicated by the results
in the  $W^\pm Z(M_T^{cut})$ row of Table~\ref{lhcrates},
if we concentrate on the transverse mass
region of   0.8 $< M_T(WZ)<$ 1.1 TeV for a 1 TeV isovector signal,
there would be about 0.8 background events  and  about 2 signal  events,
%on top of that  as a  peak,
for an integrated luminosity of  100 fb$^{-1}$.

\item The non-resonant models (LET-CG, LET-K, Delay-K) all yield observable
excesses in the $W^\pm W^\pm $ channel.

\end{itemize}

{}From Table~\ref{lhcrates} for the $W_LW_L$ fusion signals
and the predicted background rates,
we can estimate the number of LHC years
necessary to generate a signal at the 99\% Confidence Level defined
as follows. We require $B_{\rm max}< S_{min}$, where Poisson statistics
predicts that 99\% of the time
pure background would yield $B< B_{\rm max}$, while signal plus background
would yield $S>S_{\rm min}$.  We employ Poisson statistics due to
the rather small event rates in certain channels/models.
For sufficient $S$ and $B$ event numbers that Gaussian statistics can
be employed,
our 99\% confidence level corresponds roughly to $S/\sqrt{S+B} = 4$,
{\it i.e.} a statistical significance of $4\sigma$.
The results are given in Table~\ref{lhclum}.  We
see that with a few years running at the LHC, one should be able to
observe a significant enhancement in at least one gold-plated channel.
Such an enhancement would be an important step towards revealing the
physics of electroweak symmetry breaking.

%\begin{table}[h]
\begin{table}[t]
\centering
\caption{\label{lhclum}
Number of years (if $<10$) at LHC required for a
99\% confidence level signal.}
\bigskip

\begin{tabular}{l|cccccccc}
\hline\hline
& \multicolumn{8}{c}{Model}\\
 \cline{2-9}
Channel& SM & Scalar & O($2N$) & Vec~1.0 & Vec~2.5 & LET CG &
LET K & Delay K\\
\hline
$ZZ(4\ell)$ &
1.0  & 2.5  & 3.2  & \    & \   & \    & \     & \    \\
$ZZ(2\ell2\nu)$ &
0.5  & 0.75 & 1.0  & 3.7  & 4.2 & 3.5  & 4.0   & 5.7  \\
 $W^+W^-$ &
0.75 & 1.5  & 2.5  & 8.5  & \   & 9.5  & \     & \    \\
 $W^\pm Z$ &
 \   & \    & \    & 7.5  & \   & \    & \     & \    \\
$W^\pm W^\pm $ &
4.5  & 3.0  & 4.2  & 1.5  & 1.5 & 1.2  & 1.2   & 2.2  \\
\hline\hline
\end{tabular}
\end{table}

\section{Drell-Yan Production}

At the LHC energy, further sensitivity to the isovector models is possible
through the Drell-Yan process \cite{bhr,ck}.
%To see this, let us work in a basis where $W$ and $V$ refer
%to the mass eigenstates of the $W$ and the vector resonance,
%obtained after diagonalizing the full mass matrix for the mixing Lagrangian.
Assuming that the vector resonance does not couple directly to the
quarks,  the Drell-Yan production occurs only through  $W$-$V$ mixing.
%$q\overline q' \rightarrow W^* \rightarrow V \rightarrow W^\pm Z,W^+W^-$.
The spin and color averaged amplitude  for
$q\overline q' \rightarrow W^*-V \rightarrow W^\pm_L Z_L,W^+_LW^-_L$
is:
\[
\overline{|{\cal M}|}^2=
{a^2g^4(c_v^2+c_a^2)\over 96}\cdot{tu\over (s-M_V^2)^2+M_V^2\Gamma_V^2},
\]
where $c_v=c_a=1/\sqrt{2}$ for $W^\pm Z$, and
$c_v=(1-\tan^2\theta_{\rm w})t_{3L}^{(i)}$ and
$c_a=(1-\tan^2\theta_{\rm w})[t_{3L}^{(i)}-(2Q_i\sqrt{a}M_Z/M_V)
\sin^2\theta_{\rm w}]$ for $W^+W^-$.  In the latter case,
$Q_i=+{2\over 3}$ and $t_{3L}^{(i)}={1\over 2}$ for quark flavors $i=u,c,t$,
while $Q_i=-{1\over 3}$ and $t_{3L}^{(i)}=-{1\over 2}$ for $i=d,s,b$.
The constant $a$ is determined by
the width and mass of the vector resonance,
$a=192\pi v^2\Gamma_{V}/M_{V}^3$, where $v=246$ GeV.

If the isovector mass is not too large,
%and if jet-tagging is eliminated,
the signal from this process may have more statistical significance than
that for longitudinal $W^\pm Z$ scattering via $WW$ fusion
discussed above.
%with jet-tagging.
However, we must eliminate the jet tag because the mixing mechanism
does not have an accompanying spectator jet at lowest order.

In what follows we present our
result from a Born-level calculation for the signal processes, assuming
a 100\% efficiency for the central-jet veto.  This is a reasonable
approximation because for $q\bar q' \to WW$ processes, the Born-level
result is very close to that of the QCD corrected zero-jet process after
vetoing \cite{ohnemus,zerojet}.  (Note that the mixing processes do not
contribute significantly to our single-tag event rates in Table~\ref{lhcrates}.
This is because the tagging efficiencies for the mixing processes would be
1.4\% and 1.2\% in the two channels, respectively, as computed for the
continuum $WW$ cross sections.)

In Figure~\ref{rhofig} we show the transverse mass distributions for
the sum of the signals and SM backgrounds for $M_V=1$ TeV and 2.5 TeV.
Despite the increase in background that results from eliminating
the jet tag, the increase in the signal for a 1 TeV vector resonance
presents a clear bump in the $M_T$-spectrum near the resonance mass.
As expected, the $W^\pm Z$ channel via $W$-$V$ mixing  could be  best for
studying a vector resonance at the LHC if its mass is near 1 TeV.

Indeed, from Table~\ref{lhcratesii} we see that in the bin
$0.85<M_T(WZ)<1.05$ TeV,
the mixing signal has a statistical significance of $S/\sqrt B\sim
15$, far better than obtained in any of the channels after single-tagging
the spectator jets; see Table~\ref{lhcrates}.  However, for a 2.5 TeV
vector state the signal rate is too low to be observable.

The $W^+W^-$ channel seems to be less useful for observing a vector
resonance signal arising via the mixing process.  This is a consequence
of the more stringent leptonic cuts that are necessary to suppress
the larger SM backgrounds (especially the $t\bar t$ background),
as well as the lack of distinctive
peak structure in the  $M(\ell\ell)$ distribution.
The  event rates for signals and backgrounds
are shown in Table~\ref{lhcratesii}.

%\begin{table}[h]
\begin{table}[t]
\centering
\caption{\label{lhcratesii}
Event rates per LHC-year for $q\overline q\rightarrow W^+W^-$ and $q\overline
q' \rightarrow W^\pm Z$ channels, from $W$-$V$ mixing and backgrounds, compared
to the corresponding  $W_LW_L$ fusion signal rates, after removing the
jet-tagging
cut in Table~\protect\ref{cutstable}, assuming $\protect\sqrt s=14~{\rm TeV}$,
an annual luminosity of $100 ~{\rm fb}^{-1}$, and $m_t=175~{\rm GeV}$.}
\bigskip

\begin{tabular}{l|ccc}
\hline\hline
 & Bkgd. & Vec~1.0: $W$-$V$ mix / fusion & Vec~2.5: $W$-$V$ mix
 /  fusion\\
\hline
$W^+W^-$ \hfill & 420 & 8.6 / 10 & 0.3 / 9.0 \\
%$M_{\ell\ell} > 0.5$ \hfill &  330 & 6.7 / 8.9 & 0.2 / 7.9 \cr
\hline
$W^\pm Z$ \hfill & 220 & 73 / 8.7 & 1.4 / 6.4 \\
\hline
$W^\pm Z$ \hfill & & $0.85<M_T(WZ)<1.05$ TeV     &  $2<M_T(WZ)<2.8$ TeV  \\
Bkgd./mix/fusion &   & 22/ 69  / 3.2 &   0.82/0.81/0.55\\
%$W^\pm Z$ \hfill & & $0.8<M_T(WZ)<1.1$ TeV     &  $2<M_T(WZ)<2.8$ TeV  \\
%Bkgd./mix/fusion &   & 34/ 71  / 4.4 &   0.82/0.81/0.55\\
\hline\hline
\end{tabular}
\end{table}

\section{Discussion}

Having presented our results, we will close with a few comments.

\begin{enumerate}

\item[(a)]
A systematic comparison of the different gold-plated modes
allows one to distinguish
between the different models to a certain degree.
Models with a scalar isospin-zero resonance (SM, Scalar and O($2N$))
will yield a large excess of events in the
$ZZ\to 2\ell2\nu$, $ZZ \to 4\ell$, and $W^+W^-\to \ell^+\ell^-\nu\nu$
final states, a feature that is very distinct from predictions
of the other models; those with a vector isospin-one resonance
with $M_V \sim$ 1 TeV can be studied
most easily in the $W^\pm Z \to \ell^\pm \nu2\ell$ channel via $W$-$V$ mixing;
while models with heavier vector resonances or no resonances at all imply a
large enhancement in the $W^\pm W^\pm \to \ell^\pm \ell^\pm 2\nu$ channels.

\item[(b)]
As mentioned previously, we have used EWA/ET techniques to calculate the
signal cross sections for all the models.  We have checked that for a 1
TeV Higgs boson, and using our cuts, the EWA/ET results agree with the
exact SM calculation to about 10\% for the non-resonant channels ($W^\pm Z,
W^\pm W^\pm$) and to about a factor of two for those with an $s$-channel
resonance ($ZZ, W^+ W^-$).  The discrepancy is caused by difficulties in
treating the large resonant width in the ET formalism.  It should be
emphasized, however, that the EWA/ET techniques give accurate results
for non-resonant channels only in the large $M_{WW}$ and central rapidity
regions. On the other hand, we have carried out  exact calculations for  the
$W_TW_T$ and $W_TW_L$ background processes at ${\cal O}(\alpha^4)$;
for these processes EWA calculations are not applicable \cite{ewa}.

\item[(c)]
For our signal calculations of $W^{}_L W^{}_L$ scattering, we have
chosen the QCD scale in the parton distribution functions
to be  $M^{}_W$ and have ignored higher order
QCD effects.  It is shown \cite{qcdww} that with such a scale choice,
the QCD corrections to the $W^{}_L W^{}_L$ scattering processes
are negligible and our signal results are thus of rather small theoretical
uncertainty. Moreover, since we have concentrated on the exclusive channels
with a tagged jet and (most of the time) vetoing central jets, effects
on our analyses from  additional QCD radiation are small,
for both signal and background processes.

\item[(d)]
We have found that tagging a single energetic forward jet is
effective in suppressing the large backgrounds, especially those
from continuum $WW$ production, including  $W^\pm Z$ as a background
to $W^\pm W^\pm $.  Based on results from detector simulations at high
luminosities \cite{sdc,atlas,cms}, we have taken the jet-tagging threshold
to be $p_T(j)>$ 40 GeV.  Whether it is more advantageous to tag both of the
forward jets, or just one, depends crucially on the minimum $p_T(j)$
threshold below which tagging is not feasible.  The $W_LW_L$ scattering
signal tends to give jets with quite small $p_T$  ($\langle p_T\rangle\sim
M_W/2$), so that many of the spectator jets have $p_T$'s below 40 GeV.
This means that if the minimum $p_T$ threshold is set high
($\mathrel{\raise.2ex\hbox{$>$}\hskip-.8em\lower.9ex\hbox{$\sim$}} 40~{\rm
GeV}$), then only single-tagging gives a reasonable efficiency for retaining
the signal events of interest.  On the other hand, if a much lower $p_T$
threshold can be used, then double-tagging is quite efficient for isolating
the $W_LW_L$ signal.
In Ref.~\cite{atlas} it is claimed that the pile-up background is quite small
after {\it double}-tagging, even for a threshold as low as $15~{\rm GeV}$.
If such a low threshold for double-tagging can be employed, one can
consider relaxing some of the leptonic cuts to obtain larger signal
rates. Note however, that the threshold of 40 GeV we applied refers to a
value at the parton level; in contrast, the threshold of 15 GeV of
Ref.~\cite{atlas} is imposed after full fragmentation and jet reconstruction,
and may correspond to a higher parton-level value.  Fragmentation will,
of course, spread out the momenta of the spectator jets. We have not
attempted to model the effects of fragmentation. However, our
parton-level threshold is sufficiently high that it should be easy
to implement an appropriate corresponding (presumably lower)
threshold for reconstructed spectator jets after fragmentation.
This corresponding threshold should be determined using detailed
detector Monte Carlos (that also incorporate calorimeter losses
and the like) by the ATLAS and CMS detector groups.

Further study will be needed to determine whether isolation of
the $W_LW_L$ signal is better accomplished via single-tagging
with strong lepton cuts ~\cite{baggeretali}
or double-tagging with weaker lepton cuts \cite{atlas}.
In practice, it will undoubtedly prove fruitful to analyze the data
following both procedures.  However, we wish to emphasize that
single-tagging with strong leptonic cuts
has the advantage of minimizing the influence of uncertainties associated
with $W_TW_T,W_TW_L$  final states.  This is because double-tagging with
weak leptonic cuts, although  yielding  a larger signal event rate,
has a much smaller ratio of the $W_LW_L$ to the
$W_TW_T +W_TW_L$ background. Thus, a 20\% change or uncertainty in the
transversely polarized $W$ background  would be
confused with the $W_LW_L$ signal in the case of double-tagging with
weak lepton cuts, but not in the single-tagging/strong-lepton-cuts
analysis. Uncertainties in the background rate predictions
will inevitably be present. In particular, systematic errors in
the Monte Carlo simulation predictions for the background levels
in the various channels will arise due to uncertainties
in parton distribution functions, detector response, higher-order corrections
and so forth. In addition, by analyzing the data with strong cuts
we are able to separate the effects of new physics contributions
to the $W_LW_L$ sector from those that might or might not be
present in the $W_TW_T,W_TW_L$ sectors.

\item[(e)]
For the isovector models considered, we have altered our parameter
choices from those considered in Ref.~\cite{baggeretali}.  The larger
separation in mass (as compared to our earlier choices of 2.0 and 2.5
TeV) provides a better comparison between resonances with high and moderate
masses.  Also, the new widths are significantly smaller than those in our
previous work.  The new widths are roughly the largest allowed that are
consistent with the current experimental limits on mixing \cite{besslimit}
between the $Z$ and the neutral vector boson $V^0$ for the models
considered.
It is important to note that since cross
sections for processes with $s$-channel vector resonances are roughly
proportional to the width, our signal rates are rather small.  Were the
width constraints relaxed, as possible in other models,
the signal rates could be substantially  enhanced \cite{ck}.

\item[(f)]
For  the non-resonant models, we have  mainly concentrated
on the leading-order universal term in a chiral Lagrangian \cite{eff},
namely, the LET amplitude.  Beyond the leading order,
we have considered one special  case in which
the scattering amplitudes are computed to second order in the energy
expansion and the counterterms are chosen to delay the unitarity violation
to energies beyond $2~{\rm TeV}$ \cite{bdv}.  The motivation is to
provide a conservative scenario for the non-resonant model.
More generally, one should include any
higher-order effects and other non-linear operators
that are present in the chiral Lagrangian; the specific forms
for such terms reflect the underlying dynamics of a strongly interacting
electroweak symmetry breaking sector \cite{etconf}.

\end{enumerate}

Comparing to the results of Ref.~\cite{baggeretali}, we see that lowering the
LHC energy to $\sqrt s=14~{\rm TeV}$ weakens the purely leptonic gold-plated
signals for a strongly interacting electroweak symmetry breaking sector.
Nonetheless, by careful optimization of the cuts, we find that an observable
excess of events can be seen for all of the strongly interacting models that
we consider, after several years of running  with an annual
luminosity of  $100~{\rm fb}^{-1}$.
Increased rates and significance would be possible if a much lower threshold
for jet tagging were possible, despite the large number of pile-up events
accompanying the $W_LW_L$ scattering reactions and their backgrounds.
We emphasize that the search for strong electroweak symmetry breaking
requires, in large part, detecting the signal as only an overall enhancement
in rates above the SM backgrounds.  Therefore the systematics of the
experimental measurements must be fully under control.

\section*{Acknowledgments}

We would like to thank J.~Ohnemus for providing us with
the FORTRAN codes to evaluate the processes $q\bar q' \to WW$
with 1-loop QCD corrections; and G. Azuelos for discussions
regarding Ref.~\cite{atlas}.
The work of J. Bagger was supported in part by NSF grant
PHY-9404057. The work of V.~Barger was supported
in part by DOE grant DE-FG02-95ER40896.
The work of K. Cheung was supported in part by DOE-FG03-93ER40757.
The work of J. Gunion and T. Han was supported in part by DOE grant
DE-FG-03-91ER40674.
The work of G. Ladinsky was supported in part by
NSF Grant PYH-9209276.
The work of R. Rosenfeld was supported by CNPq in Brazil.
C.--P. Yuan was supported in part by NSF grant PHY-9309902.
Further support was provided to J. Gunion and T. Han by the Davis Institute
for High Energy Physics and to V. Barger by the Wisconsin Alumni Research
Foundation.

\newpage
\section*{Figure Captions}

\begin{enumerate}

\item\label{sewsfigi}  Invariant mass distributions for the
gold-plated purely leptonic final states that arise
from the processes $pp \rightarrow ZZX \to 4\ell X$,
$pp \rightarrow ZZX \to 2\ell  2\nu X$,
$pp \rightarrow W^+W^-X$,
$pp \rightarrow W^\pm ZX$ and
$pp \rightarrow W^\pm W^\pm X$,
for $\sqrt s = 14$ TeV and an annual LHC luminosity of 100~fb$^{-1}$.
The signal is plotted above the summed background.  The mass variable of
the $x$-axis is in units of GeV and the bin size is 50 GeV.  Distributions are
presented for:
(a) the SM with a 1 TeV Higgs boson;
(b) the O($2N$) model with $N=2$ and cutoff $\Lambda = 3$ TeV;
(c) a chirally coupled scalar with $M_S = 1$ TeV, $\Gamma_S = 350$ GeV;
(d) a chirally coupled vector with $M_V = 1$ TeV, $\Gamma_V = 5.7$ GeV;
(e) a chirally coupled vector with $M_V = 2.5$ TeV, $\Gamma_V = 520$ GeV;
(f) the LET-CG non-resonant model unitarized following Chanowitz and Gaillard;
(g) the LET-K non-resonant model unitarized by the $K$-matrix prescription;
(h) the Delay-K ${\cal O}(p^4)$ non-resonant model,
unitarized by the K-matrix prescription.

\item\label{rhofig}  Transverse mass distributions for
$pp \rightarrow W^* - V \to W^\pm ZX$ signals
for (a) $M_V=1$ TeV, $\Gamma_V = 5.7$ GeV
and (b) $M_V=2.5$ TeV, $\Gamma_V = 520$ GeV.
The  signal  is plotted above the
summed SM background. The mass variable of the
$x$-axis is in units of GeV  and the bin size is 50 GeV.

\end{enumerate}
\end{document}